\documentclass[12pt]{article}
\usepackage{epsf}
\usepackage{amsmath,amssymb}
\usepackage{pdfpages}
\usepackage{graphicx}
\usepackage{comment}
\usepackage{tikz}
\usetikzlibrary{positioning}
\usepackage{physics}

\begin{document}

\newcommand{\rem}[1]{{$\spadesuit$\bf #1$\spadesuit$}}

\renewcommand{\thefootnote}{\fnsymbol{footnote}}
\setcounter{footnote}{0}

\begin{titlepage}

\def\thefootnote{\fnsymbol{footnote}}

\begin{center}
\hfill {\tt STUPP-21-244}\\

\vskip .1in

\hfill April, 2021\\

\vskip .75in

{\Large \bf

  Leptophilic Gauge Bosons \\[2mm]
  at ILC Beam Dump Experiment

}

\vskip .5in

{\large
  Kento Asai$^{(a,b)}$, Takeo Moroi$^{(a)}$ and Atsuya Niki$^{(a)}$
}

\vskip 0.5in

$^{(a)}$
{\em
Department of Physics, The University of Tokyo, Tokyo 113-0033, Japan
}

\vskip 0.2in

$^{(b)}$
{\em 
Department of Physics, Saitama University, 255 Shimo-Okubo, Sakura-ku,\\
Saitama 338-8570, Japan
}

\end{center}
\vskip .5in

\begin{abstract}

  We study the prospects of searching for leptophilic gauge bosons
  (LGBs) at the beam dump experiment using $e^\pm$ beams of
  International $e^+e^-$ Linear Collider (ILC).  We consider LGBs in
  association of $U(1)_{e-\mu}$, $U(1)_{e-\tau}$, and
  $U(1)_{\mu-\tau}$ gauge symmetries, which are assumed to be light
  and long-lived.  Utilizing the energetic electron and positron beams
  of the ILC, we show that the ILC beam dump experiment can cover the
  parameter regions which have not been explored before.  We also
  discuss the possibility of distinguishing various models.

\end{abstract}

\end{titlepage}

\renewcommand{\thepage}{\arabic{page}}
\setcounter{page}{1}
\renewcommand{\thefootnote}{\#\arabic{footnote}}
\setcounter{footnote}{0}
\renewcommand{\theequation}{\thesection.\arabic{equation}}

\section{Introduction}
\label{sec:intro}
\setcounter{equation}{0}

The ILC \cite{Behnke:2013xla, Baer:2013cma, Adolphsen:2013jya,
  Adolphsen:2013kya, Behnke:2013lya} is one of the prominent
possibilities of high energy collider experiments to study physics
beyond the standard model (BSM).  Because of the high energy $e^\pm$
beams as well as precise understandings of elementary processes and
backgrounds, it has been discussed that $e^+e^-$ collisions may
provide us rich information.  On the contrary, it was unclear if the
ILC can play important roles in the search of very weakly interacting
particles, which are hardly produced by the $e^+e^-$ collision, even
though there are many candidates of such particles in BSM models.

In Ref.~\cite{Kanemura:2015cxa}, it has been pointed out that the ILC
can be also a facility to study very weakly interacting particles
utilizing high energy $e^\pm$ beams if a detector can be installed
behind the beam dump.  The proposal was to install a shield, a veto,
and a decay volume behind the beam dump of the ILC to detect and to
study light and long-lived BSM particles which can be produced by the
scattering of the electron and positron with the material in the dump.
There are several advantages of the beam dump experiment at the ILC.
First, very high energy electrons and positrons are available.
Second, because all the electrons and positrons will be dumped after
each collision at the ILC experiment, an enormous number of the
electrons and positrons can be used for the beam dump experiment.  For
the hidden photon (i.e., hidden $U(1)$ gauge boson which couples to
the standard model particles only through the kinetic mixing with the
photon)~\cite{Holdom:1985ag} and axion-like particles (for
reviews,~\cite{Jaeckel:2010ni,Ringwald:2012hr,Ringwald:2012cu}), it
has been shown that the beam dump experiment at the ILC can cover the
parameter regions which have not been explored yet
\cite{Kanemura:2015cxa, Sakaki:2020mqb}.  There exist, however, other
potential targets of the ILC beam dump experiment.

In order to understand what kind of information we can obtain with the
beam dump experiment at the ILC, we consider models with leptophilic
gauge bosons (LGBs) and discuss the possibility to study the LGBs at
the ILC beam dump experiment.  In models with LGBs, we introduce a new
$U(1)$ gauge symmetry under which only some of the leptons are
charged.  In particular, here we consider flavor dependent $U(1)$
gauge symmetries associated with the difference of lepton flavors
$U(1)_{e-\mu}$, $U(1)_{e-\tau}$, and
$U(1)_{\mu-\tau}$~\cite{Foot:1990mn,He:1990pn,He:1991qd,Foot:1994vd}.  These
gauge symmetries can be added to the standard model without the gauge
anomaly, and the various previous studies have discussed their effects
on the phenomenology, such as the neutrino
oscillation~\cite{Bell:2000vh, Joshipura:2003jh, Bandyopadhyay:2006uh,
  Samanta:2010zh}, neutrino mass matrix~\cite{Araki:2012ip,
  Heeck:2014sna, Asai:2017ryy, Asai:2018ocx, Asai:2019ciz}, and so on.
In particular, the $U(1)_{\mu-\tau}$ can contribute to muon anomalous
magnetic moment~\cite{Ma:2001tb,Baek:2001kca}.  In addition, it can
alleviate the Hubble tension~\cite{Escudero:2019gzq, Araki:2021xdk};
we will see that the ILC beam dump experiment may cover some parameter
region where such an alleviation is realized.

Compared to the dark photon, the LGB has direct couplings to leptons
so (i) the LGB production can be dominated by the direct electron-LGB
coupling in the $U(1)_{e-\mu}$ and $U(1)_{e-\tau}$ models, and (ii)
the LGB decays into neutrino pair (as well as into charged lepton
pair) which may significantly reduce the number of signal events.
Taking into account these features, we study the prospect of the study
of the LGBs at the ILC beam dump experiment.  We also discuss the
possibility to distinguish the LGBs and the dark photon after the
discovery using the fact that the LGB decays dominantly into leptons
while the dark photon decays into both charged leptons and hadrons.

The organization of this paper is as follows.  In Section
\ref{sec:production}, we summarize the model with the LGB and explain
how the expected number of signal events is estimated.  In Section
\ref{sec:reach}, we show the sensitivity of the ILC beam dump
experiment to the LGBs.  In Section \ref{sec:discrimination}, we
consider how and how well various models of LGBs (and the dark photon)
can be distinguished by using the measurements of the decay modes.
Section \ref{sec:conclusions} is devoted to conclusions and
discussion.

\section{Production of Leptophilic Gauge Boson}
\label{sec:production}
\setcounter{equation}{0}

In this section, we explain the procedure to calculate the number of
signals of the LGB at the ILC beam dump experiment.  Here, we assume a
basic setup of the beam dump system; following
Refs.\ \cite{Adolphsen:2013jya, Adolphsen:2013kya}, we consider the
case that the electron and positron beams after the collision are
dumped into beam dumps (with the length of $L_{\rm dump}$) filled with
$\rm {H_2O}$.  With the injection of the electron (or position) beam
into the dump, the LGB (denoted as $X$) can be produced by the
scattering process $e^\pm N \rightarrow e^\pm X N'$ (with $N$ and $N'$
being nuclei).  As discussed in the previous section, we assume that
the decay volume with detectors (with the length of $L_{\rm dec}$) as
well as the muon shield between the beam dump and the decay volume
(with the length of $L_{\rm sh}$) are installed.  Then, once produced,
the LGB flies (because it is boosted) and decays into a lepton pair.
If the decay happens inside the decay volume (with the length of
$L_{\rm dec}$), it can be observed as a signal of the production of
BSM particles.

For the calculation of the event rate, the relevant part of the
Lagrangian is given as follows:
\begin{align}
  {\cal L} = &\,
  -\frac{1}{4} A_{\mu\nu} A^{\mu\nu}  
  -\frac{1}{4} X_{\mu\nu} X^{\mu\nu}
  -\frac{\epsilon_0}{2} A_{\mu\nu} X^{\mu\nu}
  + \frac{1}{2} m_X^2 X_\mu X^\mu
  \nonumber \\ &\, +
  \sum_{\ell=e, \mu, \tau}
  \left[
    \bar{\ell}
    \left\{ \gamma^\mu
    \left( \partial_\mu - i e_{\rm EM} A_\mu -i g\, Q_\ell X_\mu \right)
    - m_\ell
    \right\} \ell
    + 
    \bar{\nu}_\ell
    \gamma^\mu
    \left( \partial_\mu -i g\, Q_f X_\mu \right) P_L
    \nu_\ell
    \right],
\end{align}
where $A_{\mu\nu}$ is the field strength of $U(1)_{\rm EM}$ gauge
boson while $X_{\mu\nu}$ is that of the LGB.  We consider the cases of
$(Q_e,Q_\mu,Q_\tau)=(1,-1,0)$, $(1,0,-1)$, or $(0,1,-1)$, which correspond to 
$U(1)_{e-\mu}$, $U(1)_{e-\tau}$, and $U(1)_{\mu-\tau}$ models, respectively.

The LGB mixes with the photon via the ``tree-level'' mixing parameter
$\epsilon_0$ as well via loop diagrams with charged leptons inside the
loop. (We call $\epsilon_0$ as the ``tree-level'' mixing parameter
even though it may be generated by the loop effects of BSM particles; see
the discussion below.)  We define the effective mixing parameter as
\begin{align}
  \epsilon_{\rm eff} \equiv |\epsilon_0 + \Delta \epsilon (q^2=m_X^2)|,
\end{align}
where
\begin{align}
  \Delta \epsilon (q^2) \equiv
  - \frac{e_{\rm EM} g}{2\pi^2} \sum_{\ell=e, \mu, \tau} Q_\ell
  \int_0^1 dx x (1-x)
  \log \left[ m_\ell^2 - q^2 x (1-x) - i 0 \right],
\end{align}
with $q$ being the four-momentum of the LGB.  When $\sqrt{q^2}$ is
much smaller than the masses of charged leptons, $|\Delta
\epsilon|/g\simeq 0.027$, $0.042$, and $0.014$ for the $U(1)_{e-\mu}$,
$U(1)_{e-\tau}$, and $U(1)_{\mu-\tau}$ models, respectively.  It is
often the case that the tree-level mixing parameter $\epsilon_0$ is
assumed to be zero.  However, $\epsilon_0$ depends on ultraviolet
physics.  It can be non-vanishing at the cut-off scale (which may be
as large as the Planck scale).  Furthermore, even if it vanishes at
the cut-off scale, $\epsilon_0$ can be generated if there exist scalar
bosons or fermions which are charged under the leptophilic $U(1)$ and
$U(1)_{\rm EM}$ (or, above the electroweak scale, $U(1)_Y$).  For
simplicity, we parameterize
\begin{align}
  \epsilon_{\rm eff}
  = \kappa_\epsilon |\Delta \epsilon (q^2=m_X^2)|.
\end{align}
For the case of $\epsilon_0=0$, $\kappa_\epsilon=1$.  As we see below,
the number of the LGB events at the ILC beam dump experiment is
sensitive to the value of $\kappa_\epsilon$ in particular in the
$U(1)_{\mu-\tau}$ model.

The LGB decays via the tree-level coupling with leptons and via the
mixing with the photon.  The partial decay rates to the charged
leptons are calculated as
\begin{align}
  \Gamma (X\rightarrow \ell^+ \ell^-) = 
  \frac{g_\ell^2}{12\pi} m_X \left( 1 + \frac{2m_\ell^2}{m_X^2} \right)
  \sqrt{1 - \frac{4m_\ell^2}{m_X^2}},
\end{align}
where the effective $\ell$-$\ell$-$X$ coupling constant is defined
as\footnote
{We expect that $\epsilon_{\rm eff}$ is one-loop suppressed and that
  $e_{\rm EM} |\epsilon_{\rm eff}|\ll g$.  Thus, for the case of
  $Q_\ell=\pm 1$, we neglect the effect of the LGB-photon mixing.}
\begin{align}
  g_\ell
  \equiv
  \left\{ \begin{array}{ll}
    g & :~ Q_\ell = \pm1 , \\[1mm]
    e_{\rm EM} |\epsilon_{\rm eff}| & :~ Q_\ell=0 ,
  \end{array} \right.
\end{align}
while those to neutrinos are found to be
\begin{align}
  \Gamma (X\rightarrow \nu_\ell \bar{\nu}_\ell) = 
  \frac{g^2 Q_\ell^2}{24\pi} m_X.
\end{align}
The total decay rate of $X$ is obtained as
\begin{align}
  \Gamma_X \equiv
  \sum_{\ell=e, \mu, \tau}
  \left[ 
    \Gamma (X\rightarrow \ell^+ \ell^-) +
    \Gamma (X\rightarrow \nu_\ell \bar{\nu}_\ell)
    \right].
\end{align}
It is notable that the LGBs decay into neutrino pairs as well as to
charged lepton pairs and that the branching ratios of the neutrino
final states are always sizable.  In particular, in the
$U(1)_{\mu-\tau}$ model, the LGB decays dominantly into the neutrino
pairs when $m_X<2m_\mu$; this makes it difficult to search for the
$U(1)_{\mu-\tau}$ gauge boson at the ILC beam dump experiment, as
shown in the following.

Now, we discuss the calculation of the event rate. We follow the
procedure adopted in Ref.\ \cite{Kanemura:2015cxa}, where the possibility of
searching for dark photon at the ILC beam dump experiment was studied.
For the scattering process with the nucleus with the atomic number $A$
and charge $Z$, the expected number of signals is given by
\begin{align}
  N_{\rm sig} = N_e \frac{N_{\rm Avo} X_0}{A}
  B_{\rm sig}
  \int_{m_X}^{E_{\rm beam}-m_e} dE_X \int_{E_X+m_e}^{E_{\rm beam}} dE_e 
  \int_0^T dt 
  \frac{I_e(E_{\rm beam},E_e,t)}{E_e}
  \left. \frac{d\sigma}{dx} \right|_{x=E_X/E_e}
  P_{\rm dec},
  \label{Nsig}
\end{align}
where $N_e$ is the total number of electrons injected into the dump,
$N_{\rm Avo}$ is the Avogadro constant, $X_0\simeq
716.4A/[Z(Z+1)\ln (287/Z^{1/2})]\ {\rm g/cm^2}$ is the radiation length,
$T\equiv \rho L_{\rm dump}/X_0$ with $\rho$ being the density of
water, and 
$I_e$ is the  energy distribution of $e^\pm$ after passing through a
medium of the radiation length $t$ \cite{Tsai:1986tx}:
\begin{align}
  I_e(E_{\rm beam},E_e,t) = \frac{1}{E_{\rm beam}}
  \frac{[\ln (E_{\rm beam}/E_e)]^{bt - 1}}{\Gamma (bt)},
\end{align}
with $b=\frac{4}{3}$.  The branching ratio of $X$ into the signal
mode, $B_{\rm sig}$, is taken to be
\begin{align}
  B_{\rm sig} = 
  \frac{1}{\Gamma_X}
  \sum_{\ell=e, \mu, \tau}
  \Gamma (X\rightarrow \ell^+ \ell^-).
\end{align}
Assuming that $L_{\rm dump}$ is long enough so that the $X$ production
mostly occurs near the edge of the dump, the probability of the decay
of $X$ in the decay volume, $P_{\rm dec}$, is approximated as
\begin{align}
  P_{\rm dec} = 
  e^{-\gamma_X^{-1} \Gamma_X (L_{\rm dump}+L_{\rm sh})}
  (1-e^{-\gamma_X^{-1} \Gamma_X L_{\rm dec}}),
\end{align}
where $\gamma_X\equiv E_X/m_X$.  The differential cross section for
the elementary process, $e^\pm N\rightarrow e^\pm X N'$, is calculated
by using the Weizs\"acker-Williams approximation \cite{Kim:1973he,
  Bjorken:2009mm, Andreas:2012mt}:
\begin{align}
  \frac{d \sigma}{dx}
  = 
  \frac{\alpha^2 g_{e}^2}{\pi}
  \chi
  \left( 1 - x + \frac{1}{3}x^2 \right)
  \left( \frac{1-x}{x} m_X^2 + x m_e^2 \right) ^{-1}
  \sqrt{1 - \frac{m_X^2}{E_e^2}},
\end{align}
where $\chi$ is the effective flux of photons (see
Ref.\ \cite{Kanemura:2015cxa}), and $\alpha$ is the fine structure constant.

\section{Discovery Reach}
\label{sec:reach}
\setcounter{equation}{0}

Now we calculate the expected number of signal events.  We adopt the
ILC experiment and its beam dump system proposed
\cite{Adolphsen:2013jya, Adolphsen:2013kya} as well as the setup of
the beam dump experiment discussed in Ref.\ \cite{Kanemura:2015cxa}:
\begin{itemize}
\item The beam dump consists of high-pressure (10 bar) water vessel.
  The length of the dump is $L_{\rm dump}=11\ {\rm m}$ ($30X_0$).
  Thus, in our calculation, the target is ${\rm H_2 O}$.
\item We assume that a shield is installed behind the dump in order to
  remove the standard model backgrounds, in particular, those from
  muons.  One of the possibilities is to install a carefully designed
  magnetic field as proposed in the SHiP experiment \cite{Anelli:2015pba}.
  One may also consider a thick lead shield \cite{Sakaki:2020mqb}.  The
  detail of the shield is beyond the scope of our analysis; we assume
  that a well designed shield is installed so that the standard model
  backgrounds become negligible.  In the following, we concentrate on
  the production process of the LGB which is independent of the detail
  of the shield.
\item We assume a decay volume with particle detectors behind the
  shield.  The length of the decay volume is assumed to be $L_{\rm
    dec}=50\, {\rm m}$.
\item The ILC bunch train contains 1312 bunches, each of which has
  $2\times 10^{10}$ electrons.  The frequency of the dump of the bunch
  train is $5\ {\rm Hz}$.  Thus, with one-year operation, about
  $4\times 10^{21}$ electrons and positrons are injected into the
  dump.
\end{itemize}

\begin{figure}
  \centering
  \includegraphics[width=0.68\linewidth]{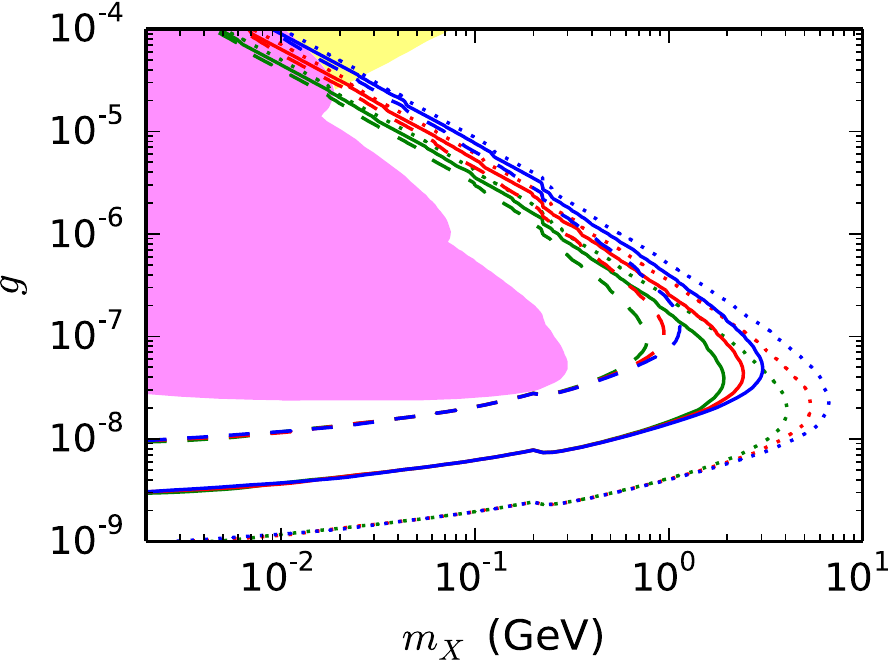}
  \caption{Contours of expected number of signal events for the
    $U(1)_{e-\mu}$ model.  The beam energy is taken to be $E_{\rm
      beam}=125$ (green), $250$ (red), and $500\, {\rm GeV}$ (blue).
    The dotted, solid, and dashed lines are for $N_{\rm sig}=10^{-2}$,
    $1$, and $10^2$, respectively, taking $N_e=4\times 10^{21}$.  The
    mixing parameter is taken to be $\kappa_\epsilon=1$.  The pink and
    yellow shaded regions are excluded by beam dump and
    neutrino-electron scattering experiments, respectively.}
  \label{fig:nev110}
  \vspace{5mm}
  \includegraphics[width=0.7\linewidth]{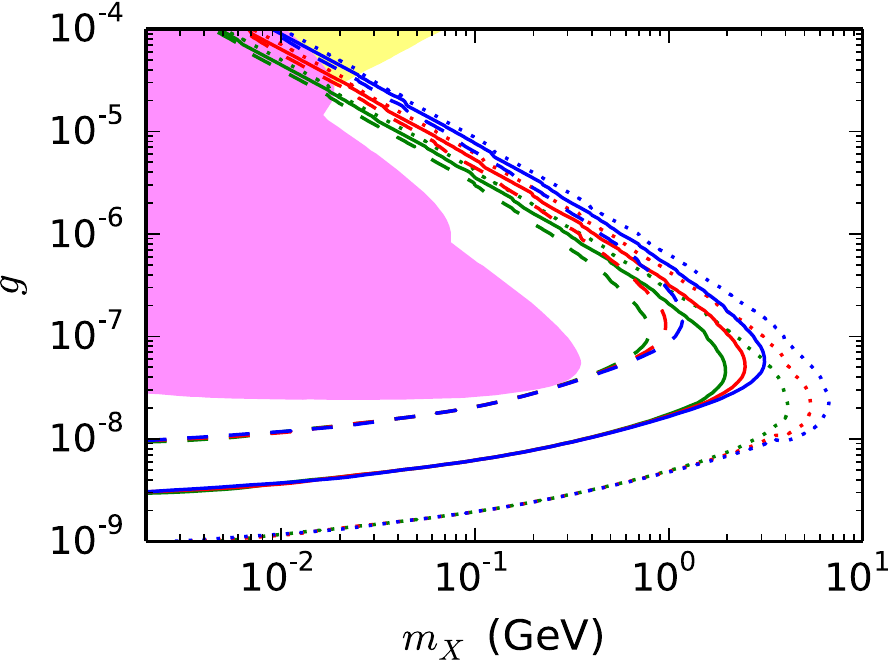}
  \caption{Same as Fig.\ \ref{fig:nev110}, but for the $U(1)_{e-\tau}$
    model.}
  \label{fig:nev101}
\end{figure}

We first consider the case that electron is charged under the
leptophilic $U(1)$ gauge interaction.  In Figs.\ \ref{fig:nev110} and
\ref{fig:nev101}, we show the expected number of signal events $N_{\rm
  sig}$ for the cases of $U(1)_{e-\mu}$ and $U(1)_{e-\tau}$ models on
$m_X$ vs.\ $g$ plane.  The dotted, solid, and dashed lines correspond
to $N_{\rm sig} = 10^{-2}, 1$, and $10^2$, respectively, for the beam
energy taken to be $E_{\rm beam} = 125$ (green), $250$ (red), and
$500$ GeV (blue).  The experimental bounds are also drawn; the pink
and yellow shaded regions show the bounds from beam dump experiments
in the past and the $\bar{\nu}_e$-$e^-$
scattering~\cite{Bauer:2018onh}, respectively.  Requiring $N_{\rm
  sig}\sim 1$ for the discovery, we can see that the ILC beam dump
experiment may cover the parameter region which has not been excluded
yet.  We can see that the ILC beam dump experiment does not have a
sensitivity to the cases with $g$ being too small or too large.  The
insensitivity to the case of small $g$ can be understood from the fact
that the production cross section and the probability of the decay
inside the decay volume are both suppressed in the limit of
$g\rightarrow 0$.  On the contrary, with very large value of $g$, the
decay rate of the LGB becomes so enhanced that most of the LGBs
produced at the dump decay before reaching the decay volume.

The discovery reaches for the $U(1)_{e-\mu}$ and $U(1)_{e-\tau}$
models are similar; this is because the productions of the LGBs in
these models occur due to the direct coupling between the LGBs and the
electron or positron, and also that the branching fractions of the
LGBs into charged lepton pairs are significant because they can decay
into $e^+e^-$ pair (in particular when $m_X<2m_\mu$).  The dominant
(visible) decay modes of these LGBs are, however, significantly
different.  For the case of $U(1)_{e-\mu}$, dominant visible decay
modes of the LGB are $X\rightarrow e^+e^-$ and $\mu^+\mu^-$, while the
LGB of the $U(1)_{e-\tau}$ model decays mainly into $e^+e^-$ (if
$m_X<2m_\tau$) as well as neutrino pairs.  These features may be used
to distinguish the model behind the LGB, as we will discuss in the
next section.

\begin{figure}[t]
  \centering
  \includegraphics[width=0.68\linewidth]{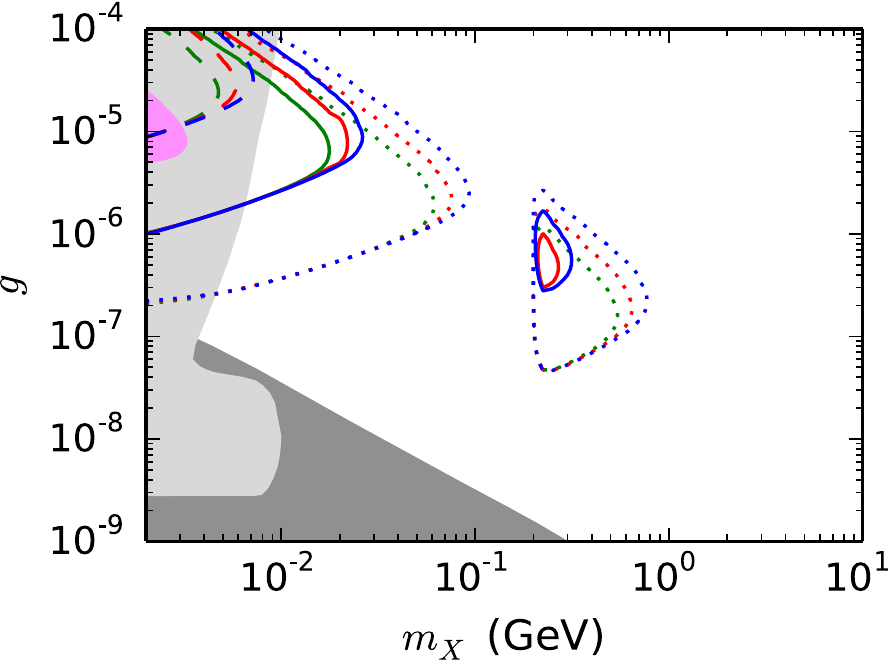}
  \caption{Contours of expected number of signal events for the
    $U(1)_{e-\tau}$ model.  The beam energy is taken to be $E_{\rm
      beam}=125$ (green), $250$ (red), and $500\, {\rm GeV}$ (blue).
    The dotted, solid, and dashed lines are for $N_{\rm sig}=10^{-2}$,
    $1$, and $10^2$, respectively, taking $N_e=4\times 10^{21}$.  The
    mixing parameter is taken to be $\kappa_\epsilon=1$.  The pink
    region is excluded by beam dump experiment. The light- and
    dark-gray regions show the constraint from the BBN and SN1987A,
    respectively.}
  \label{fig:nev011}
\end{figure}

\begin{figure}[t]
  \centering
  \includegraphics[width=0.68\linewidth]{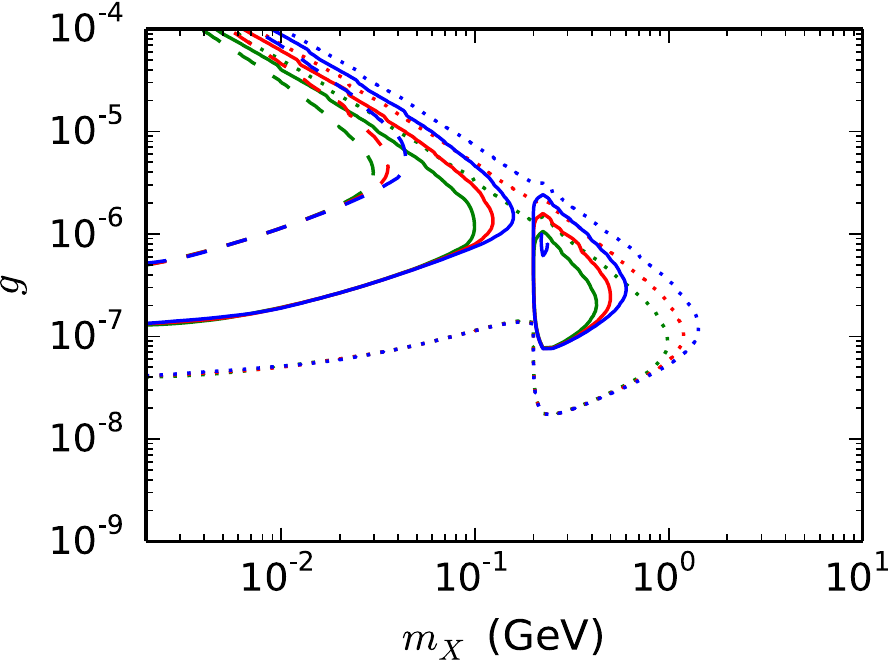}
  \caption{Same as Fig.\ \ref{fig:nev110}, but for the
    $U(1)_{\mu-\tau}$ model with $\kappa_\epsilon=5$.}
  \label{fig:nev011k5}
\end{figure}

Next, we consider the case of the $U(1)_{\mu-\tau}$ model; in
Fig.\ \ref{fig:nev011}, the expected number of signal events is shown,
taking $\kappa_\epsilon=1$.  The constraints from the beam dump
experiment~\cite{Bauer:2018onh}, big bang nucleosynthesis
(BBN)~\cite{Escudero:2019gzq}, and SN1987A~\cite{Croon:2020lrf} are
shown in the pink, light gray, and dark gray shaded regions,
respectively.  We can see that the ILC beam dump experiment may access
the parameter region which has not been excluded yet.  In particular,
the region with $(m_X,g)\sim (10\, {\rm MeV}, 10^{-5})$ may be
covered; such a parameter region is interesting because the Hubble
tension may be alleviated \cite{Escudero:2019gzq}.  As one can see,
compared to the cases of the $U(1)_{e-\mu}$ and $U(1)_{e-\tau}$
models, the expected number of the signal events is significantly
suppressed.  This is because, in the $U(1)_{\mu-\tau}$ model, the LGB
does not directly couples to the electron (or positron) which is the
incident particle to cause the LGB production.  The LGB production is
via the LGB-photon mixing and hence is significantly suppressed
because $e_{\rm EM} |\epsilon_{\rm eff}|\ll g$.  We can also see that
$N_{\rm sig}$ is enhanced when $m_X> 2m_\mu$.  This is due to the fact
that, in the $U(1)_{\mu-\tau}$ model, the LGB directly couples to
$\mu^\pm$ and hence Br$(X\rightarrow\mu^+\mu^-)$ is sizable if the
process is kinematically allowed.  Thus, when $m_X> 2m_\mu$, we can
use $\mu^+\mu^-$ final state for the LGB search while, for $m_X<
2m_\mu$, we can only use $X\rightarrow e^+e^-$ final state whose
branching fraction is very small in the $U(1)_{\mu-\tau}$ model.

Because the production of the LGB is though the LGB-photon mixing in
the $U(1)_{\mu-\tau}$ model, the production cross section of the LGB
is sensitive to the tree-level mixing parameter $\epsilon_0$ (and
hence to $\epsilon_{\rm eff}$).  In addition, the mixing is important
for the decay process $X\rightarrow e^+e^-$ which only provides
the visible final state when $m_X<2m_\mu$.  Consequently, $N_{\rm sig}$ is
highly dependent on $\epsilon_{\rm eff}$.  In particular, if
$\epsilon_{\rm eff}$ is enhanced, the ILC beam dump experiment has a
higher sensitivity to the parameter space (i.e., $m_X$ vs.\ $g$ plane)
of the $U(1)_{\mu-\tau}$ model.  To see this, 
In Fig.\ \ref{fig:nev011k5}, 
we show the contours of
constant $N_{\rm sig}$, taking $\kappa_\epsilon=5$.\footnote
{In Fig.\ \ref{fig:nev011k5}, the constraints from the beam dump
  experiment, BBN and SN1987A, are not shown because we could not find
  the constraints for the case with the tree-level mixing parameter.}
We can see that, for fixed values of $m_X$ and $g$, the expected
number of the signal events is enhanced with the increase of
$\kappa_\epsilon$.

\section{Discrimination of Models}
\label{sec:discrimination}
\setcounter{equation}{0}

Even though a signal of the LGB production can be found at the ILC
beam dump experiment, it is unclear how and how well the properties of
the newly found particle (i.e., in the present case, the LGB) can be
studied after the discovery.  In this section, we consider the
situation
after the discovery of the LGB at the ILC beam dump experiment
and discuss whether we may understand
its properties.

The information about the LGB mass is embedded in the momentum
information of the final state particles.  In particular, the LGB may
decay into $e^+e^-$ and/or $\mu^+\mu^-$ pair.  Thus, if the momenta
(or the energy) of the visible final state particles are measured with
a reasonable accuracy, the mass of the LGB can be determined.

With identifying the final state particles from the decay, information
about the decay modes of $X$ can be also obtained.  Such a study is of
interest in particular when $m_X\gtrsim 2m_\mu$ because the possible
final states as well as the branching ratios to those final states are
strongly dependent on the model.  If $m_X< 2m_\mu$, on the contrary,
visible decay mode is only $X\rightarrow e^+e^-$ irrespective of the
model.  For such a light LGB, its properties are hardly clarified by
the study of the decay modes.  Hereafter, we assume that, after the
observations of the signals, the mass of $X$ is determined with some
accuracy and is found to be $2m_\mu<m_X<2m_\tau$.  (As shown in the
previous section, the ILC beam dump experiment may not have a good
sensitivity to LGBs heavier than $\sim 2m_\tau$.)  

In the following, we discuss the discrimination of the models based on
the decay modes of $X$.  We pay attention to the early stage of the
discovery at which $\sim 10$ signal events are available to study the
properties of the LGB.  In order to estimate how well we can determine
the decay properties of the LGB, we consider a simple analysis using
the following quantities:
\begin{align}
  r_e \equiv &\, \frac{\Gamma (X\rightarrow e^+ e^-)}
  {\Gamma_X^{({\rm vis})}},
  \\
  r_\mu \equiv &\, \frac{\Gamma (X\rightarrow \mu^+ \mu^-)}
  {\Gamma_X^{({\rm vis})}},
  \\
  r_h \equiv &\, \frac{\Gamma (X\rightarrow {\rm hadrons})}
  {\Gamma_X^{({\rm vis})}},
\end{align}
where $\Gamma_X^{({\rm vis})}$ is the partial width with visible final
states. In the case of $m_X<2m_\tau$, $\Gamma_X^{({\rm vis})}$ becomes
the following:
\begin{align}
  \Gamma_X^{({\rm vis})} \equiv 
  \Gamma (X\rightarrow e^+ e^-) + \Gamma (X\rightarrow \mu^+ \mu^-)
  + \Gamma (X\rightarrow {\rm hadrons}).
\end{align}
We consider the case of $m_X<2m_\tau$ so that $r_e+r_\mu+r_h=1$.\footnote
{When $m_X>2m_\tau$, the LGB may decay into $\tau^+\tau^-$ pair.
  Because $\tau^\pm$ can decay leptonically and hadronically, the
  final state has more variety.  However, the analysis proposed in the
  following can be applied even when $m_X>2m_\tau$ with defining $r_h$
  as $r_h\equiv 1-r_e-r_\mu$. For the case of $m_X>2m_\tau$,
  Eqs.\ \eqref{r(e-mu)} -- \eqref{r(darkph)} are modified.}
For
the case of the LGB of the $U(1)_{e-\mu}$ gauge symmetry,
\begin{align}
  r_e^{(e-\mu)} \simeq r_\mu^{(e-\mu)} \simeq 0.5,~~~
  r_h^{(e-\mu)} \simeq 0,
  \label{r(e-mu)}
\end{align}
while for $U(1)_{e-\tau}$, 
\begin{align}
  r_e^{(e-\tau)} \simeq 1.0,~~~
  r_\mu^{(e-\tau)} \simeq r_h^{(e-\tau)} \simeq 0.
\end{align}
Notice that one interesting question is whether the LGBs may be
distinguished from the dark photon (denoted as $\gamma'$) which
universally couples to all the charged particles via the kinetic
mixing with the electromagnetic photon.  For the case of dark photon
(with $m_\mu\ll m_{\gamma'}\ll m_\tau$), we find
\begin{align}
  r_e^{(\gamma')} \simeq r_\mu^{(\gamma')} \simeq \frac{1}{2+R}\ ,~~~
  r_h^{(\gamma')} \simeq \frac{R}{2+R}\ ,
  \label{r(darkph)}
\end{align}
where $R$ is the $R$-ratio for $q^2=m_{\gamma'}^2$, with $m_{\gamma'}$
being the mass of the dark photon.  This may help to distinguish the
LGBs and the dark photon because the LGBs dominantly decay into
leptonic final states.

In order to study how well various models can be distinguished after
the discovery of the signal events, we assume a situation in which the
signal events with $e^+e^-$, $\mu^+\mu^-$, and hadronic final states
are observed with the event numbers of $n_e^{\rm (obs)}$, $n_\mu^{\rm
  (obs)}$, and $n_h^{\rm (obs)}$, respectively.  Then, treating $r_e$
and $r_\mu$, as well as the total number of signal events, as model
parameters, we adopt the following likelihood function on the $r_e$
vs.\ $r_\mu$ plane:
\begin{align}
  \mathcal{L} (r_e, r_\mu, r_h; n_e^{\rm (obs)}, n_\mu^{\rm (obs)}, n_h^{\rm (obs)})
  = \mathcal{N}
  \mathcal{P} (r_e n_{\rm tot}; n_e^{\rm (obs)})
  \mathcal{P} (r_\mu n_{\rm tot}; n_\mu^{\rm (obs)})
  \mathcal{P} (r_h n_{\rm tot}; n_h^{\rm (obs)}),
\end{align}
where $r_h=1-r_e-r_\mu$, $\mathcal{N}$ is the normalization factor,
$n_{\rm tot}\equiv n_e^{\rm (obs)}+n_\mu^{\rm (obs)}+n_h^{\rm (obs)}$,
and $\mathcal{P}(n,n^{\rm (obs)})\equiv n^{n^{\rm (obs)}}
e^{-n}/n^{\rm (obs)}!$ is the Poisson distribution function.\footnote
{In the parameter space of the model, one might define the likelihood
  function on the $r_e$ vs.\ $r_\mu$ plane by a projection along the
  axis of the total number of signal events:
  \begin{align*}
    \tilde{\mathcal{L}}
    (r_e, r_\mu, r_h; n_e^{\rm (obs)}, n_\mu^{\rm (obs)}, n_h^{\rm (obs)})
    \equiv \tilde{\mathcal{N}}
    \int dn \mathcal{J} (n)
    \mathcal{P} (r_e n, n_e^{\rm (obs)})
    \mathcal{P} (r_\mu n, n_\mu^{\rm (obs)})
    \mathcal{P} (r_h n, n_h^{\rm (obs)}),
  \end{align*}
  where $\tilde{\mathcal{N}}$ is the normalization factor and
  $\mathcal{J}(n)$ is the measure of the integration. It can be easily
  seen that $\tilde{\mathcal{L}}=\mathcal{L}$ irrespective of
  $\mathcal{J}$ because $r_e+r_\mu+r_h=1$.}
The normalization factor $\mathcal{N}$ is determined so that 
\begin{align}
  \int_0^1 dr_e \int_0^{1-r_e} dr_\mu
  \mathcal{L} (r_e, r_\mu, r_h; n_e^{\rm (obs)}, n_\mu^{\rm (obs)}, n_h^{\rm (obs)})
  = 1.
\end{align}
Using the above likelihood function, we determine the allowed regions
with given confidence levels.

\begin{figure}
  \centering
  \includegraphics[width=0.65\linewidth]{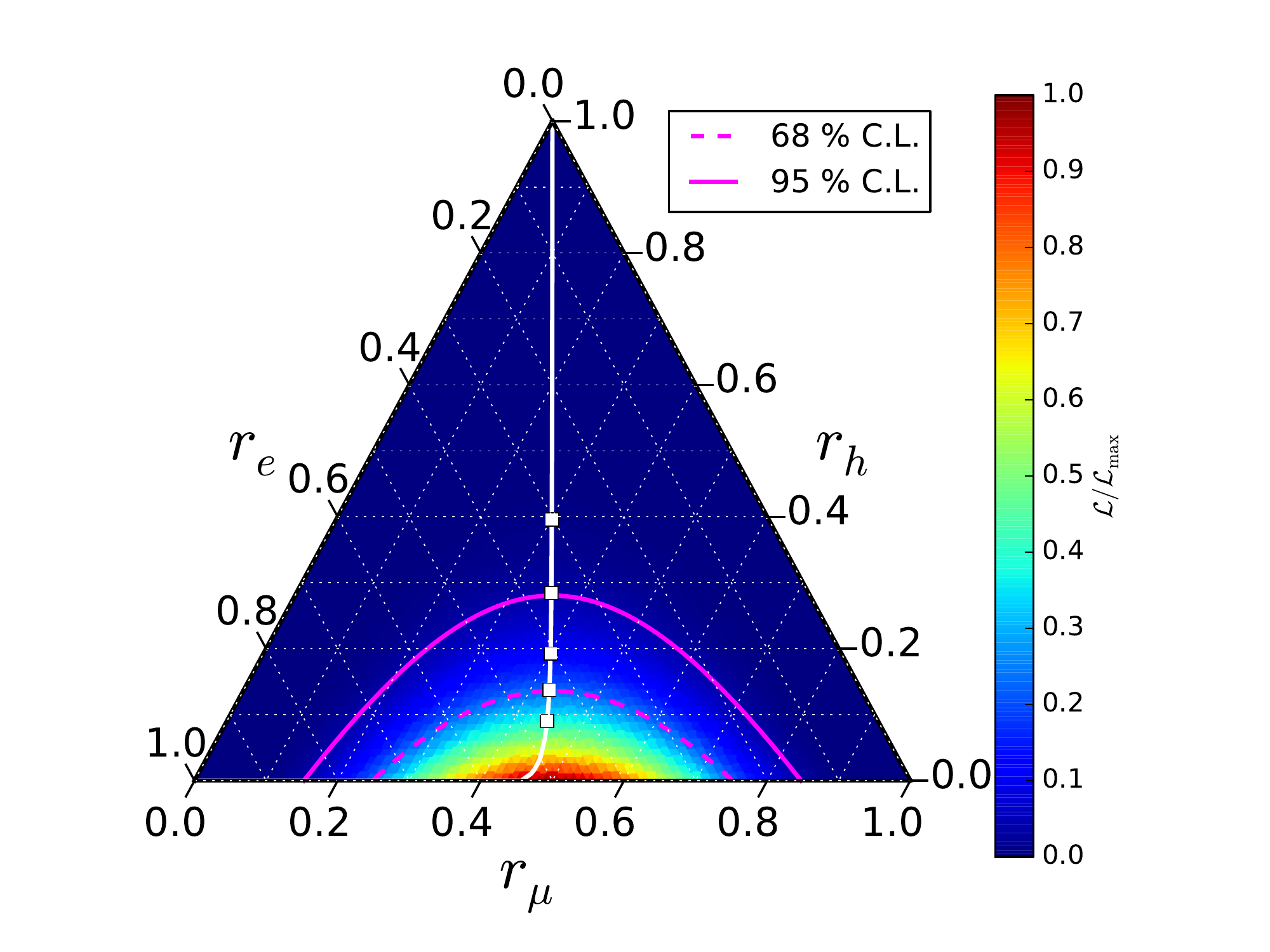}
  \caption{Expected constraint on $r_e$, $r_\mu$, and $r_h$ parameters
    in the $U(1)_{e-\mu}$ model for $(n_e^{\rm (obs)}, n_\mu^{\rm
      (obs)}, n_h^{\rm (obs)})=(5,5,0)$.  The magenta lines are
    expected constraint at $68\, \%$ (dashed) and $95\, \%$ (solid)
    C.L.  The white line shows the prediction of the dark photon
    model; squares on the white line are the predictions for the cases
    of $m_{\gamma'}=400$, $450$, $500$, $550$, and $600\, {\rm MeV}$
    from below.  The background color indicates the ratio
    $\mathcal{L}/\mathcal{L}_{\rm max}$, where $\mathcal{L}_{\rm max}$
    is the maximal value of the likelihood.}
  \label{fig:emu_10events}
  \vspace{5mm}
  \includegraphics[width=0.65\linewidth]{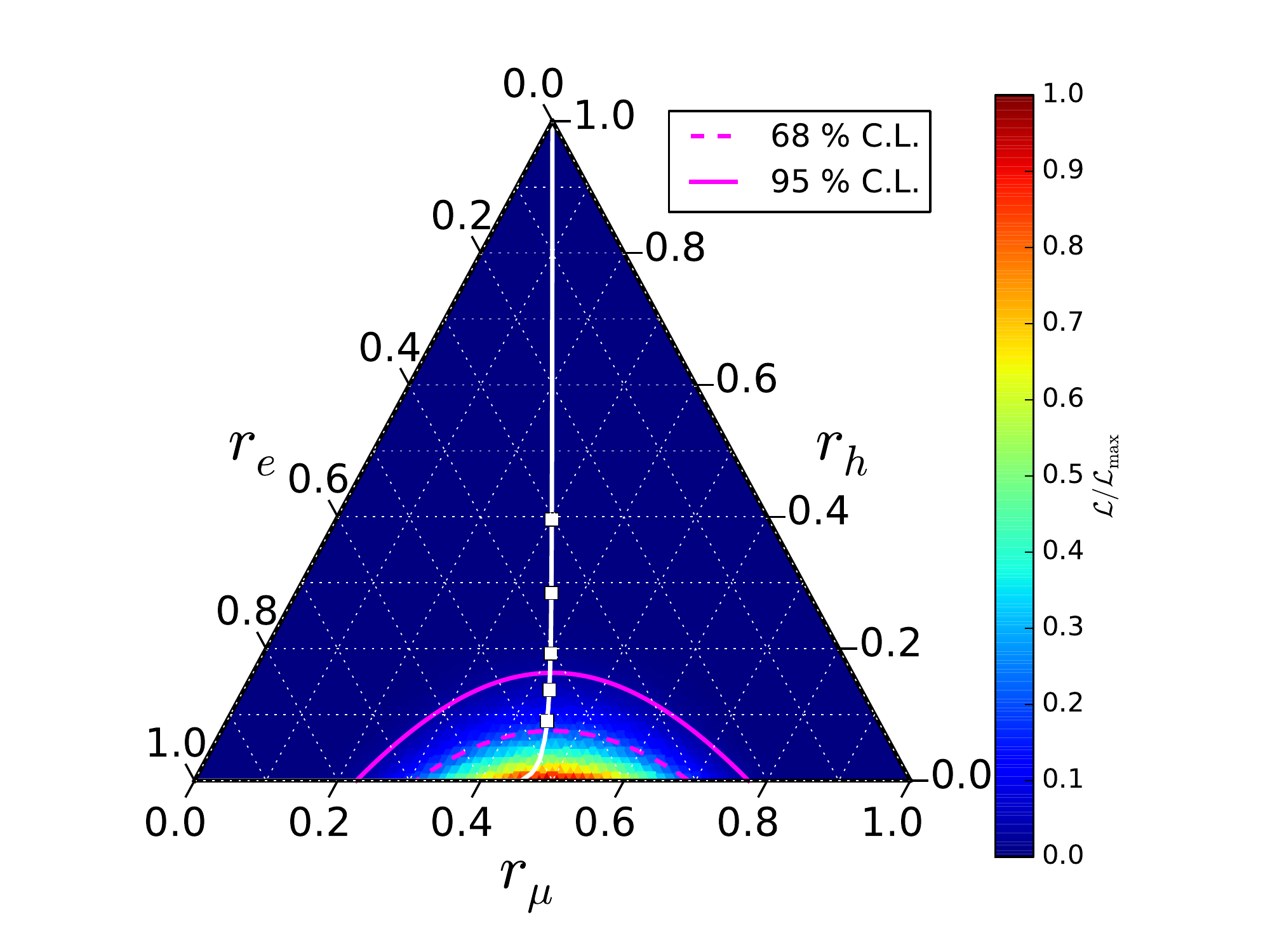}
  \caption{Same as Fig.\ \ref{fig:emu_10events}, but for 
    $(n_e^{\rm (obs)}, n_\mu^{\rm (obs)}, n_h^{\rm (obs)})=(10,10,0)$.}
  \label{fig:emu_20events}
\end{figure}

First, we consider the $U(1)_{e-\mu}$ LGB.  In
Figs.\ \ref{fig:emu_10events} and \ref{fig:emu_20events}, the expected
constraint on $r_e$, $r_\mu$ and $r_h$ are given on the ternary plot,
taking $(n_e^{\rm (obs)}, n_\mu^{\rm (obs)}, n_h^{\rm (obs)})=(5,5,0)$
and $(10,10,0)$, respectively.  With the discovery of $\sim 10$ --
$20$ signals of the $U(1)_{e-\mu}$ LGB, we may discriminate the
possibility of the $U(1)_{e-\tau}$ model which predicts
$(r_e,r_\mu,r_h)\simeq (1,0,0)$.  On the figures, we also show the
prediction of the dark photon, which is given as a line because the
prediction of the dark photon depends on the $R$-ratio.  For the case
of the dark photon, $r_h^{(\gamma')}$ can take any value between $0$
and $\sim 1$ because $R$ varies from $\sim 0$ to $\sim 50$
\cite{Zyla:2020zbs}; the predictions for the cases of
$m_{\gamma'}=400$, $450$, $500$, $550$, and $600\, {\rm MeV}$ are also
plotted.  For the case of the $U(1)_{e-\mu}$ model, the parameter
region of the dark photon model with $r_h^{(\gamma')}\sim 0$ is hardly
excluded solely by the study of the decay mode; if information about
the mass of the LGB becomes available, the dark photon model may be
excluded.  For example, $r_h^{(\gamma')}\gtrsim 0.3$ if
$m_{\gamma'}\gtrsim 550\, {\rm MeV}$.  Thus, if $m_X$ is found to be
heavier than $\sim 550\, {\rm MeV}$, we may exclude the possibility of
the dark photon with accumulating $\sim 20$ signal events from the
decay of the $U(1)_{e-\mu}$ LGB.

\begin{figure}
  \centering
  \includegraphics[width=0.65\linewidth]{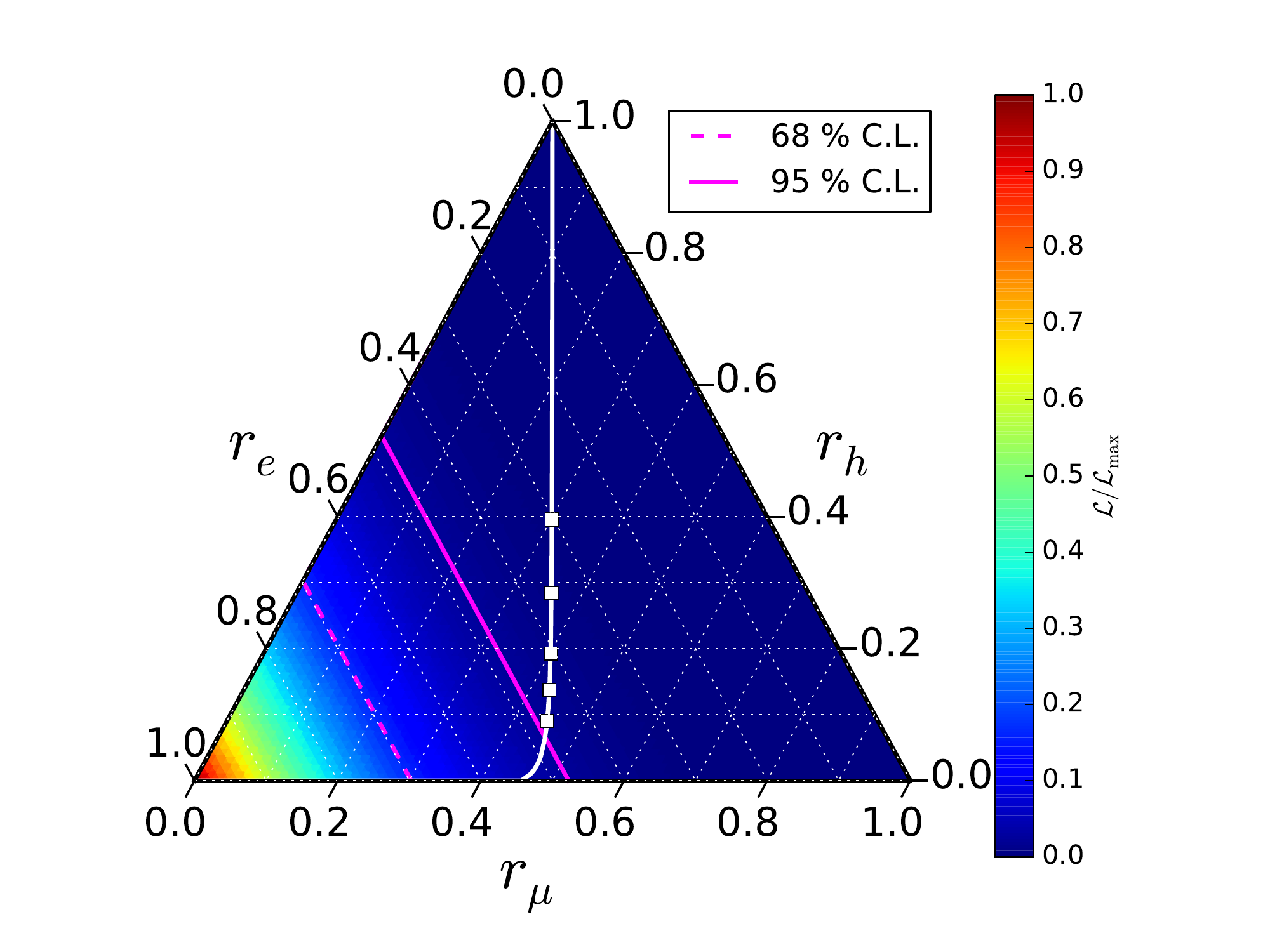}
  \caption{Same as Fig.\ \ref{fig:emu_10events}, but for 
    the $U(1)_{e-\tau}$ model taking $(n_e^{\rm (obs)}, n_\mu^{\rm
      (obs)}, n_h^{\rm (obs)})=(5,0,0)$.}
  \label{fig:etau_05events}
  \vspace{5mm}
  \includegraphics[width=0.65\linewidth]{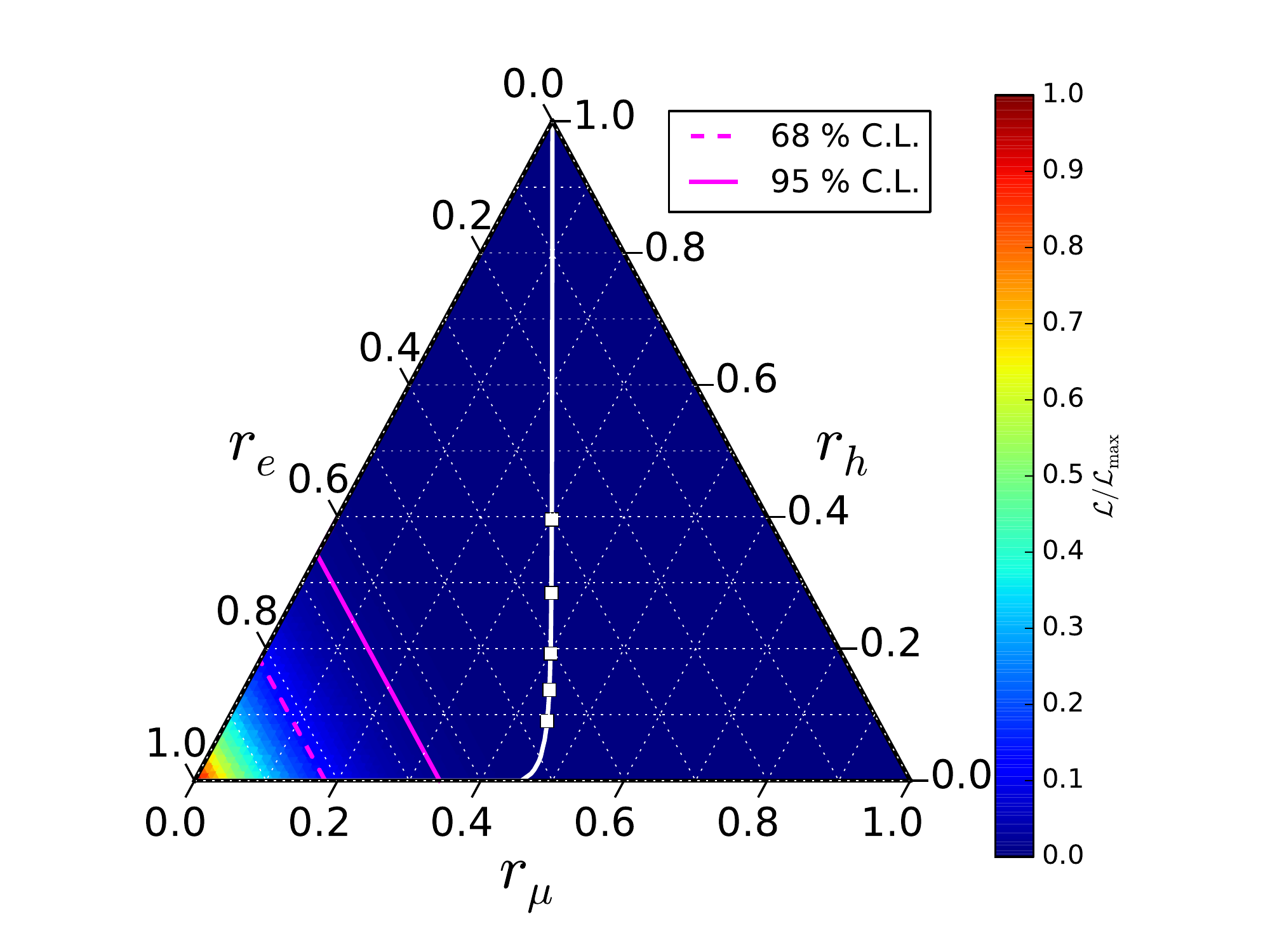}
  \caption{Same as Fig.\ \ref{fig:emu_10events}, but for 
    the $U(1)_{e-\tau}$ model taking
    $(n_e^{\rm (obs)}, n_\mu^{\rm (obs)}, n_h^{\rm (obs)})=(10,0,0)$.}
  \label{fig:etau_10events}
\end{figure}

For the case of the $U(1)_{e-\tau}$ model, the LGB dominantly decays
into the $e^+e^-$ final state (as well as into the neutrino pair); the
effect of the mixing is negligibly small when the number of signal
events is $\sim 10$.  In Figs.\ \ref{fig:etau_05events} and
\ref{fig:etau_10events}, we show the expected constraint on $r_e$,
$r_\mu$ and $r_h$ parameters, taking $(n_e^{\rm (obs)}, n_\mu^{\rm
  (obs)}, n_h^{\rm (obs)})=(5,0,0)$ and $(10,0,0)$, respectively.  We
can see that the LGB of the $U(1)_{e-\tau}$ model can be distinguished
from that of the $U(1)_{e-\mu}$ model and the dark photon if more than
$\sim 5$ signal events are observed, assuming that the mass of the LGB
is measured to be heavier than $\sim 2m_\mu$.

So far, we did not consider using the number of signal events $N_{\rm
  sig}$ for the study of the LGB properties. As indicated by
the figures, $N_{\rm sig}$ is sensitive to the mass and the coupling
constants so that the measurement of the event rate may provide
information about the mass and the coupling constant of the LGB.  In
particular, the event rate may be sizable in the $U(1)_{e-\mu}$ and
$U(1)_{e-\tau}$ models even if the LGB mass is relatively large, while
it is relatively suppressed in the $U(1)_{\mu-\tau}$ model even if the
LGB mass is small.  Thus, combining with the kinematical determination
of the LGB mass, we may distinguish the $U(1)_{\mu-\tau}$ model from
others using the event rate information.  We should, however, note
that the production cross section of the LGB, as well as the
efficiency of the signal detection, should be well understood for the
study using the event rate; such a study is beyond the scope of our
analysis and we leave it as a future task.

\section{Conclusions and Discussion}
\label{sec:conclusions}
\setcounter{equation}{0}

In this paper, we have considered the possibility to search for the
LGBs, i.e., gauge bosons in association with $U(1)_{e-\mu}$,
$U(1)_{e-\tau}$, and $U(1)_{\mu-\tau}$ gauge symmetries, at the ILC
beam dump experiment.  Installing shields, vetos, and decay volume
with detectors, the ILC beam dump experiment is capable of detecting
the signals of the LGBs with very small couplings which are hardly
studied by the conventional ILC setup only with the detector for the
energetic $e^+e^-$ collision.\footnote
{If the leptophilic gauge coupling is sizable, the LGB may be searched
  with $e^+e^-$ colliders \cite{Araki:2017wyg, Jho:2019cxq}.}
Assuming that the LGBs are light and
weakly interacting, we have calculated the expected number of events
and confirmed that a significant number of LGB signals may be
possible.  We have seen that the ILC beam dump experiment can cover
the parameter regions which have not been explored before.  Such a
study may open a new window to BSM physics.

In our calculation, we have concentrated on the production process
$e^\pm N \rightarrow e^\pm X N'$, which exists irrespective of the
species of the injected beam and the detailed setup of the shield.  In
fact, there may exist other production processes of the LGBs for a
particular setup.  For example, if a thick lead is adopted as the muon
shield, the high intensity muons produced at the dump may also play
the role of the initial-state particle to produce the LGBs
\cite{Sakaki:2020mqb}.  Such a possibility may be of interest
particularly for the case of $U(1)_{\mu-\tau}$ model for which the ILC
beam dump experiment has less sensitivity than the $U(1)_{e-\mu}$ and
$U(1)_{e-\tau}$ models if we just consider the production process of
$e^\pm N \rightarrow e^\pm X N'$.  In addition, if the positron is
used as the initial-state particle injected into the dump, the pair
annihilation process $e^+e^-\rightarrow X$ (with $e^-$ being the
electron in ${\rm H_2O}$) may also contribute to the production
process \cite{Sakaki_LCWS2021}.  These production processes require
particular setups of the experiment, and have not been studied in our
analysis; the effects of these processes on the search of the LGBs
will be discussed elsewhere \cite{AsaiMoroiNiki_Future}.

\vspace{2mm}
\noindent{\it Acknowledgments:} 
This work was supported by JSPS KAKENHI Grant Numbers JP19J13812 (KA),
16H06490 (TM), and 18K03608 (TM).  All the ternary plots are generated
with {\tt python-ternary}~\cite{pythonternary}.

%%%%%%%%%%%%%%%%%%%%%%%%%%%%%%%%%%%%%%%

\bibliographystyle{jhep}
\bibliography{ref}

\providecommand{\href}[2]{#2}\begingroup\raggedright\begin{thebibliography}{10}

\bibitem{Behnke:2013xla}
T.~Behnke et~al., eds., \emph{{The International Linear Collider Technical
  Design Report - Volume 1: Executive Summary}},
  \href{https://arxiv.org/abs/1306.6327}{{\ttfamily 1306.6327}}.

\bibitem{Baer:2013cma}
H.~Baer et~al., eds., \emph{{The International Linear Collider Technical Design
  Report - Volume 2: Physics}},
  \href{https://arxiv.org/abs/1306.6352}{{\ttfamily 1306.6352}}.

\bibitem{Adolphsen:2013jya}
C.~Adolphsen et~al., eds., \emph{{The International Linear Collider Technical
  Design Report - Volume 3.I: Accelerator \& in the Technical Design Phase}},
  \href{https://arxiv.org/abs/1306.6353}{{\ttfamily 1306.6353}}.

\bibitem{Adolphsen:2013kya}
C.~Adolphsen et~al., eds., \emph{{The International Linear Collider Technical
  Design Report - Volume 3.II: Accelerator Baseline Design}},
  \href{https://arxiv.org/abs/1306.6328}{{\ttfamily 1306.6328}}.

\bibitem{Behnke:2013lya}
H.~Abramowicz et~al., \emph{{The International Linear Collider Technical Design
  Report - Volume 4: Detectors}},
  \href{https://arxiv.org/abs/1306.6329}{{\ttfamily 1306.6329}}.

\bibitem{Kanemura:2015cxa}
S.~Kanemura, T.~Moroi and T.~Tanabe, \emph{{Beam dump experiment at future
  electron\textendash{}positron colliders}},
  \href{https://doi.org/10.1016/j.physletb.2015.10.002}{\emph{Phys. Lett. B}
  {\bfseries 751} (2015) 25}
  [\href{https://arxiv.org/abs/1507.02809}{{\ttfamily 1507.02809}}].

\bibitem{Holdom:1985ag}
B.~Holdom, \emph{{Two U(1)'s and Epsilon Charge Shifts}},
  \href{https://doi.org/10.1016/0370-2693(86)91377-8}{\emph{Phys. Lett. B}
  {\bfseries 166} (1986) 196}.

\bibitem{Jaeckel:2010ni}
J.~Jaeckel and A.~Ringwald, \emph{{The Low-Energy Frontier of Particle
  Physics}},
  \href{https://doi.org/10.1146/annurev.nucl.012809.104433}{\emph{Ann. Rev.
  Nucl. Part. Sci.} {\bfseries 60} (2010) 405}
  [\href{https://arxiv.org/abs/1002.0329}{{\ttfamily 1002.0329}}].

\bibitem{Ringwald:2012hr}
A.~Ringwald, \emph{{Exploring the Role of Axions and Other WISPs in the Dark
  Universe}}, \href{https://doi.org/10.1016/j.dark.2012.10.008}{\emph{Phys.
  Dark Univ.} {\bfseries 1} (2012) 116}
  [\href{https://arxiv.org/abs/1210.5081}{{\ttfamily 1210.5081}}].

\bibitem{Ringwald:2012cu}
A.~Ringwald, \emph{{Searching for axions and ALPs from string theory}},
  \href{https://doi.org/10.1088/1742-6596/485/1/012013}{\emph{J. Phys. Conf.
  Ser.} {\bfseries 485} (2014) 012013}
  [\href{https://arxiv.org/abs/1209.2299}{{\ttfamily 1209.2299}}].

\bibitem{Sakaki:2020mqb}
Y.~Sakaki and D.~Ueda, \emph{{Searching for new light particles at the
  international linear collider main beam dump}},
  \href{https://doi.org/10.1103/PhysRevD.103.035024}{\emph{Phys. Rev. D}
  {\bfseries 103} (2021) 035024}
  [\href{https://arxiv.org/abs/2009.13790}{{\ttfamily 2009.13790}}].

\bibitem{Foot:1990mn}
R.~Foot, \emph{{New Physics From Electric Charge Quantization?}},
  \href{https://doi.org/10.1142/S0217732391000543}{\emph{Mod. Phys. Lett. A}
  {\bfseries 6} (1991) 527}.

\bibitem{He:1990pn}
X.G.~He, G.C.~Joshi, H.~Lew and R.R.~Volkas, \emph{{NEW Z-prime
  PHENOMENOLOGY}}, \href{https://doi.org/10.1103/PhysRevD.43.R22}{\emph{Phys.
  Rev. D} {\bfseries 43} (1991) 22}.

\bibitem{He:1991qd}
X.-G.~He, G.C.~Joshi, H.~Lew and R.R.~Volkas, \emph{{Simplest Z-prime model}},
  \href{https://doi.org/10.1103/PhysRevD.44.2118}{\emph{Phys. Rev. D}
  {\bfseries 44} (1991) 2118}.

\bibitem{Foot:1994vd}
R.~Foot, X.G.~He, H.~Lew and R.R.~Volkas, \emph{{Model for a light Z-prime
  boson}}, \href{https://doi.org/10.1103/PhysRevD.50.4571}{\emph{Phys. Rev. D}
  {\bfseries 50} (1994) 4571}
  [\href{https://arxiv.org/abs/hep-ph/9401250}{{\ttfamily hep-ph/9401250}}].

\bibitem{Bell:2000vh}
N.F.~Bell and R.R.~Volkas, \emph{{Bottom up model for maximal muon-neutrino -
  tau-neutrino mixing}},
  \href{https://doi.org/10.1103/PhysRevD.63.013006}{\emph{Phys. Rev. D}
  {\bfseries 63} (2001) 013006}
  [\href{https://arxiv.org/abs/hep-ph/0008177}{{\ttfamily hep-ph/0008177}}].

\bibitem{Joshipura:2003jh}
A.S.~Joshipura and S.~Mohanty, \emph{{Constraints on flavor dependent long
  range forces from atmospheric neutrino observations at super-Kamiokande}},
  \href{https://doi.org/10.1016/j.physletb.2004.01.057}{\emph{Phys. Lett. B}
  {\bfseries 584} (2004) 103}
  [\href{https://arxiv.org/abs/hep-ph/0310210}{{\ttfamily hep-ph/0310210}}].

\bibitem{Bandyopadhyay:2006uh}
A.~Bandyopadhyay, A.~Dighe and A.S.~Joshipura, \emph{{Constraints on
  flavor-dependent long range forces from solar neutrinos and KamLAND}},
  \href{https://doi.org/10.1103/PhysRevD.75.093005}{\emph{Phys. Rev. D}
  {\bfseries 75} (2007) 093005}
  [\href{https://arxiv.org/abs/hep-ph/0610263}{{\ttfamily hep-ph/0610263}}].

\bibitem{Samanta:2010zh}
A.~Samanta, \emph{{Long-range Forces : Atmospheric Neutrino Oscillation at a
  magnetized Detector}},
  \href{https://doi.org/10.1088/1475-7516/2011/09/010}{\emph{JCAP} {\bfseries
  09} (2011) 010} [\href{https://arxiv.org/abs/1001.5344}{{\ttfamily
  1001.5344}}].

\bibitem{Araki:2012ip}
T.~Araki, J.~Heeck and J.~Kubo, \emph{{Vanishing Minors in the Neutrino Mass
  Matrix from Abelian Gauge Symmetries}},
  \href{https://doi.org/10.1007/JHEP07(2012)083}{\emph{JHEP} {\bfseries 07}
  (2012) 083} [\href{https://arxiv.org/abs/1203.4951}{{\ttfamily 1203.4951}}].

\bibitem{Heeck:2014sna}
J.~Heeck, \emph{{Neutrinos and Abelian Gauge Symmetries}}, Ph.D. thesis,
  Heidelberg U., 2014.

\bibitem{Asai:2017ryy}
K.~Asai, K.~Hamaguchi and N.~Nagata, \emph{{Predictions for the neutrino
  parameters in the minimal gauged U(1)$_{L_\mu-L_\tau}$ model}},
  \href{https://doi.org/10.1140/epjc/s10052-017-5348-x}{\emph{Eur. Phys. J. C}
  {\bfseries 77} (2017) 763}
  [\href{https://arxiv.org/abs/1705.00419}{{\ttfamily 1705.00419}}].

\bibitem{Asai:2018ocx}
K.~Asai, K.~Hamaguchi, N.~Nagata, S.-Y.~Tseng and K.~Tsumura, \emph{{Minimal
  Gauged U(1)$_{L_\alpha - L_\beta}$ Models Driven into a Corner}},
  \href{https://doi.org/10.1103/PhysRevD.99.055029}{\emph{Phys. Rev. D}
  {\bfseries 99} (2019) 055029}
  [\href{https://arxiv.org/abs/1811.07571}{{\ttfamily 1811.07571}}].

\bibitem{Asai:2019ciz}
K.~Asai, \emph{{Predictions for the neutrino parameters in the minimal model
  extended by linear combination of U(1)$_{L_e-L_\mu}$, U(1)$_{L_\mu-L_\tau}$
  and U(1)$_{B-L}$ gauge symmetries}},
  \href{https://doi.org/10.1140/epjc/s10052-020-7622-6}{\emph{Eur. Phys. J. C}
  {\bfseries 80} (2020) 76} [\href{https://arxiv.org/abs/1907.04042}{{\ttfamily
  1907.04042}}].

\bibitem{Ma:2001tb}
E.~Ma and D.P.~Roy, \emph{{Anomalous neutrino interaction, muon g-2, and atomic
  parity nonconservation}},
  \href{https://doi.org/10.1103/PhysRevD.65.075021}{\emph{Phys. Rev. D}
  {\bfseries 65} (2002) 075021}
  [\href{https://arxiv.org/abs/hep-ph/0111385}{{\ttfamily hep-ph/0111385}}].

\bibitem{Baek:2001kca}
S.~Baek, N.G.~Deshpande, X.G.~He and P.~Ko, \emph{{Muon anomalous g-2 and
  gauged L(muon) - L(tau) models}},
  \href{https://doi.org/10.1103/PhysRevD.64.055006}{\emph{Phys. Rev. D}
  {\bfseries 64} (2001) 055006}
  [\href{https://arxiv.org/abs/hep-ph/0104141}{{\ttfamily hep-ph/0104141}}].

\bibitem{Escudero:2019gzq}
M.~Escudero, D.~Hooper, G.~Krnjaic and M.~Pierre, \emph{{Cosmology with A Very
  Light L$_{\mu}$ \ensuremath{-} L$_{\tau}$ Gauge Boson}},
  \href{https://doi.org/10.1007/JHEP03(2019)071}{\emph{JHEP} {\bfseries 03}
  (2019) 071} [\href{https://arxiv.org/abs/1901.02010}{{\ttfamily
  1901.02010}}].

\bibitem{Araki:2021xdk}
T.~Araki, K.~Asai, K.~Honda, R.~Kasuya, J.~Sato, T.~Shimomura et~al.,
  \emph{{Resolving the Hubble tension in a U(1)$_{L_\mu-L_\tau}$ model with
  Majoron}},  \href{https://arxiv.org/abs/2103.07167}{{\ttfamily 2103.07167}}.

\bibitem{Tsai:1986tx}
Y.-S.~Tsai, \emph{{Axion Bremsstrahlung by an Electron Beam}},
  \href{https://doi.org/10.1103/PhysRevD.34.1326}{\emph{Phys. Rev. D}
  {\bfseries 34} (1986) 1326}.

\bibitem{Kim:1973he}
K.J.~Kim and Y.-S.~Tsai, \emph{{Improved Weizsacker-Williams Method and Its
  Application to Lepton and W Boson Pair Production}},
  \href{https://doi.org/10.1103/PhysRevD.8.3109}{\emph{Phys. Rev. D} {\bfseries
  8} (1973) 3109}.

\bibitem{Bjorken:2009mm}
J.D.~Bjorken, R.~Essig, P.~Schuster and N.~Toro, \emph{{New Fixed-Target
  Experiments to Search for Dark Gauge Forces}},
  \href{https://doi.org/10.1103/PhysRevD.80.075018}{\emph{Phys. Rev. D}
  {\bfseries 80} (2009) 075018}
  [\href{https://arxiv.org/abs/0906.0580}{{\ttfamily 0906.0580}}].

\bibitem{Andreas:2012mt}
S.~Andreas, C.~Niebuhr and A.~Ringwald, \emph{{New Limits on Hidden Photons
  from Past Electron Beam Dumps}},
  \href{https://doi.org/10.1103/PhysRevD.86.095019}{\emph{Phys. Rev. D}
  {\bfseries 86} (2012) 095019}
  [\href{https://arxiv.org/abs/1209.6083}{{\ttfamily 1209.6083}}].

\bibitem{Anelli:2015pba}
{\scshape SHiP} collaboration, \emph{{A facility to Search for Hidden Particles
  (SHiP) at the CERN SPS}},  \href{https://arxiv.org/abs/1504.04956}{{\ttfamily
  1504.04956}}.

\bibitem{Bauer:2018onh}
M.~Bauer, P.~Foldenauer and J.~Jaeckel, \emph{{Hunting All the Hidden
  Photons}}, \href{https://doi.org/10.1007/JHEP07(2018)094}{\emph{JHEP}
  {\bfseries 07} (2018) 094}
  [\href{https://arxiv.org/abs/1803.05466}{{\ttfamily 1803.05466}}].

\bibitem{Croon:2020lrf}
D.~Croon, G.~Elor, R.K.~Leane and S.D.~McDermott, \emph{{Supernova Muons: New
  Constraints on $Z$' Bosons, Axions and ALPs}},
  \href{https://doi.org/10.1007/JHEP01(2021)107}{\emph{JHEP} {\bfseries 01}
  (2021) 107} [\href{https://arxiv.org/abs/2006.13942}{{\ttfamily
  2006.13942}}].

\bibitem{Zyla:2020zbs}
{\scshape Particle Data Group} collaboration, \emph{{Review of Particle
  Physics}}, \href{https://doi.org/10.1093/ptep/ptaa104}{\emph{PTEP} {\bfseries
  2020} (2020) 083C01}.

\bibitem{Araki:2017wyg}
T.~Araki, S.~Hoshino, T.~Ota, J.~Sato and T.~Shimomura, \emph{{Detecting the
  $L_{\mu}-L_{\tau}$ gauge boson at Belle II}},
  \href{https://doi.org/10.1103/PhysRevD.95.055006}{\emph{Phys. Rev. D}
  {\bfseries 95} (2017) 055006}
  [\href{https://arxiv.org/abs/1702.01497}{{\ttfamily 1702.01497}}].

\bibitem{Jho:2019cxq}
Y.~Jho, Y.~Kwon, S.C.~Park and P.-Y.~Tseng, \emph{{Search for muon-philic new
  light gauge boson at Belle II}},
  \href{https://doi.org/10.1007/JHEP10(2019)168}{\emph{JHEP} {\bfseries 10}
  (2019) 168} [\href{https://arxiv.org/abs/1904.13053}{{\ttfamily
  1904.13053}}].

\bibitem{Sakaki_LCWS2021}
Y.~Sakaki, \emph{{Talk given at LCWS2021}},
  \href{https://arxiv.org/abs/https://indico.cern.ch/event/995633/contributions/4256388}{{\ttfamily
  https://indico.cern.ch/event/995633/contributions/4256388}}.

\bibitem{AsaiMoroiNiki_Future}
K.~Asai, T.~Moroi and A.~Niki, \emph{Work in progress}.

\bibitem{pythonternary}
M.~Harper et~al., \emph{python-ternary: Ternary plots in python},
  \href{https://doi.org/10.5281/zenodo.594435}{\emph{Zenodo
  10.5281/zenodo.594435} }.

\end{thebibliography}\endgroup

%%%%%%%%%%%%%%%%%%%%%%%%%%%%%%%%%%%%%%%

\end{document}